\documentclass[preprint,showpacs,preprintnumbers,amssymb,nofootinbib]{revtex4}

\begin{document}
\title{Update on pion weak decay constants in nuclear matter }
\author{Hungchong Kim$^{1,2}$ and Makoto Oka$^{1}$
} \affiliation{$^{1}$Department of Physics,Tokyo
Institute of Technology, Tokyo 152-8552,Japan \\
$^2$ Institute of Physics and Applied Physics, Yonsei
University, Seoul 120-749, Korea.}

\begin{abstract}
The QCD sum rule calculation of the in-medium pion decay constants
using pseudoscalar-axial vector correlation function, $i \int
d^4x~ e^{ip\cdot x} \langle \rho| T[{\bar d}(x) i \gamma_5 u (x)~
{\bar u}(0) \gamma_\mu \gamma_5 d (0)] | \rho \rangle $ is
revisited. In particular, we argue that the dimension 5
condensate, $\langle {\bar q} (i D_0)^2  q \rangle_N + {1\over 8}
\langle {\bar q} g_s \sigma \cdot {\cal G} q \rangle_N$, which is
crucial for splitting the time ($f_t$) and space ($f_s$)
components of the decay constant, is not necessarily restricted to
be positive. Its positive value is found to yield a
tachyonic pion mass. Using the in-medium pion mass as an input, we
fix the dimension 5 condensate to be around $-0.025~ {\rm GeV}^2
\sim -0.019$ GeV$^2$. The role of the $N$ and $\Delta$
intermediate states in the correlation function is also
investigated. The $N$ intermediate state is found not to
contribute to the sum rules. For the $\Delta$ intermediate state,
we either treat it as a part of the continuum or propose a way to
subtract explicitly from the sum rules. With (and without)
explicit $\Delta$ subtraction while allowing the in-medium pion
mass to vary within 139 MeV $ \le m_\pi^* \le$ 159 MeV, we obtain
$f_s/f_\pi = 0.37 \sim 0.78$ and $f_t / f_\pi = 0.63 \sim 0.79$.
\end{abstract}

\maketitle

\section{INTRODUCTION}
\label{sec:intro}

The  pion decay constant in nuclear matter is one important
parameter to be determined in modern nuclear physics. As an order
parameter of spontaneous chiral symmetry breaking, its reduction
in the matter may indicate a partial restoration of chiral
symmetry. In particular, recent measurements of deeply bound
pionic
atoms~\cite{Gilg:1999qa,Itahashi:1999qb,Geissel:ur,Suzuki:2002ae}
give rise to an exciting discussion on pion-nucleus optical
potential. The isovector parameter of the s-wave pion nucleus
potential is directly related to the pion decay constant, and its
observed enhancement in nuclear matter can be interpreted as a
partial restoration of chiral
symmetry~\cite{Friedman:um,Kaiser:2001bx,Kolomeitsev:2002gc,Weise:sg},
namely by the decrease of the decay constant. This restoration
causes the universal softening of $\sigma$ and $\rho$ in the
matter~\cite{Yokokawa:2002pw}. Furthermore, the change in the
decay constant is believed to scale 
the reduction of hadron masses in the medium~\cite{Brown:kk}. Also
the decay constant is directly connected to the renormalization of
the induced pseudoscalar coupling in the matter, which is believed
to control muon captures in
nuclei~\cite{Kirchbach:1993ep,Akhmedov:ms,Kirchbach:1997rk,Akhmedov:yj}.

One interesting feature of the in-medium pion decay constant is
its separation into the time ($f_t$) and space ($f_s$) components.
For a model-independent prediction, it will
be interesting to calculate the decay constants using QCD sum
rules~\cite{Shifman:bx,Shifman:by}.
Indeed,  one of present authors (H.K) recently performed  a QCD sum rule
calculation
of the decay constants~\cite{Kim:2001xu} using pseudoscalar-axial vector
correlation function in the matter,
\begin{eqnarray}
\Pi^{\mu}=i \int d^4x~ e^{ip\cdot x} \langle \rho| T[{\bar d}(x) i
\gamma_5 u (x)~ {\bar u}(0) \gamma^\mu \gamma_5 d (0)] | \rho \rangle \ .
\label{corr}
\end{eqnarray}
This correlation function is useful because it can reproduce the
Gell-Mann$-$Oakes$-$Renner (GOR) relation in vacuum as well as its
in-medium version. Also one can clearly see the separation of $f_t$ and $f_s$ 
in QCD calculation. 
It was found that the splitting between $f_t$ and
$f_s$ is mainly driven by the dimension 5 condensate in the
nucleon, $\langle {\bar q} (i D_0)^2 q \rangle_N + {1\over 8}
\langle {\bar q} g_s \sigma \cdot {\cal G} q \rangle_N$. But its
positive value leads to a somewhat puzzling result of $f_s/f_t \ge
1$, which neither agrees with the result from in-medium chiral
perturbation theory~\cite{Kirchbach:1997rk,Meissner:2001gz},
$f_s/f_t \sim 0.28$ (or smaller), nor with the causal constraint
from Ref.~\cite{Pisarski:1996mt}, $f_s/f_t \le 1$.

However, before making a definite claim from
QCD sum rules,  one may need to re-examine the sum rule
calculation in various respects.
First, the argument leading to a positive value for the dimension 5
condensate is based on the Hermitian property
of the operators involved but, as we will discuss below, it is not
sufficient for determining
the definite sign of the dimension 5 condensate. One needs an alternative
way to constrain the  value of the dimension 5 condensate.
One possibility is to constrain it by requiring that the
sum rules reproduce a reasonable in-medium pion mass.
Another ingredient for this update is to check
more carefully the quasi-pion dominance of the
correlation function Eq.(\ref{corr}).
In particular, we need to calculate
the contributions from $N$ and $\Delta$ intermediate states
and estimate how large the change is from these intermediate states.
In this work, we address these two aspects and  improve the
previous QCD sum rule calculation for $f_t$ and $f_s$.

This paper is organized as follows.
In Sec.~\ref{sec:sumrules}, we briefly re-derive
QCD sum rules for $f_t$ and $f_s$.
The issue on the sign of the dimension 5 condensate is discussed in
Sec.\ref{sec:sign}.  The contributions from
$N$ and $\Delta$ intermediate states
are studied in Sec.~\ref{sec:nucleon} and Sec.~\ref{sec:delta}.
For the $\Delta$ contribution,
we propose a way to subtract it from our sum rules.
In Sec.~\ref{sec:analysis}, we constrain the
dimension 5 condensate by requiring it to reproduce
an acceptable in-medium pion mass within our sum rule
and use it to calculate $f_t$ and $f_s$.
We then summarize in Sec.~\ref{sec:summary}.

\section{The QCD sum rule for $f_t$ and $f_s$.}
\label{sec:sumrules}

In this section, we briefly go through the QCD sum rule
derivation of the in-medium pion decay constants, $f_t$ and $f_s$,
in Ref.~\cite{Kim:2001xu}.
Ref.~\cite{Kim:2001xu} considered the pseudoscalar-axial vector correlation
function in nuclear matter, Eq.(\ref{corr}),
and constructed the two sum rules in the following limit
\begin{eqnarray}
\Pi_t \equiv \lim_{{\bf p}\rightarrow 0}
{\Pi^0 \over i p_0}  \;; \quad \quad
\Pi_s \equiv \lim_{{\bf p}\rightarrow 0} {\Pi^j \over i p^j}\ .
\label{comb}
\end{eqnarray}
Using the in-medium decay constants defined by
\begin{eqnarray}
\langle \rho| {\bar d} \gamma^\mu \gamma_5 u |\pi^+(k) \rho \rangle
= (if_t k_0, if_s {\bf k})\ ,
\label{general}
\end{eqnarray}
and its derivative, we obtain the phenomenological sides
\begin{eqnarray}
\Pi_t^{phen}
 =
- {{m^*_\pi}^2\over 2m_q}  {{f_t}^2 \over
p^2_0-{m^*_\pi}^2 }
\;; \quad \quad
\Pi_s^{phen}
= - {{m^*_\pi}^2\over 2m_q} {f_t f_s \over
p^2_0-{m^*_\pi}^2 }\ ,
\label{phen}
\end{eqnarray}
when a quasi-pion intermediates the correlation function.
On the other hand, the operator product expansion (OPE) allows us to
calculate the correlation function using QCD degrees of freedom.
Up to dimension 5 in the expansion, the OPE side of the
correlation function is given by~\cite{Kim:2001xu},
\begin{eqnarray}
\Pi_t^{ope}
&=& -{3 \over 4\pi^2 } \int^1_0 du~ m_q
{\rm ln} [-u(1-u)p^2_0 +m_q^2 ]
+
{2 \langle {\bar q} q \rangle_\rho \over p^2_0-m_q^2}
+
{8 m_q \langle  q^\dagger i D_0 q \rangle_\rho
-2 m_q^2 \langle {\bar q} q \rangle_\rho
\over
(p^2_0 -m_q^2)^2} \ ,
\label{time}
\\
\Pi_s^{ope}
&=& -{3 \over 4\pi^2 } \int^1_0 du~ m_q
{\rm ln} [-u(1-u)p^2_0 +m_q^2 ]
+
{2\langle {\bar q} q \rangle_\rho \over p^2_0-m_q^2}
-
{8\over 3} {m_q \langle q^\dagger iD_0 q \rangle_\rho
\over
(p^2_0-m^2_q)^2}  -
{2m_q^2 \langle {\bar q} q \rangle_\rho
 \over (p^2_0-m^2_q)^2 }
\nonumber \\
&&+ {32 \over 3} \left [  \langle {\bar q} (i D_0)^2 q \rangle_\rho
+ {1\over 8}
\langle {\bar q} g_s \sigma \cdot {\cal G} q \rangle_\rho
\right ]
{1 \over (p^2_0 -m^2_q)^2} \ .
\label{space}
\end{eqnarray}
Here the subscript ``$\rho$'' denotes the nuclear expectation value.

When Eqs.(\ref{phen}),(\ref{time}) and (\ref{space})
are put into the Borel weighted sum rules,
\begin{eqnarray}
\int^{S_0}_0 ds~ e^{-s/M^2} {1 \over \pi} {\rm Im}
[\Pi_l^{phen} (s) - \Pi_l^{ope} (s)]=0  ~~~~(l=t,s) \ ,
\end{eqnarray}
we obtain the two sum rules,
\begin{eqnarray}
{{m^*_\pi}^2 \over 2 m_q} {f_t}^2 e^{-{m^*}^2_\pi/M^2}
&=&
{3 m_q \over 4 \pi^2} \int^{S_0}_{4m_q^2} d s e^{-s/M^2}
\sqrt {1 - {4 m_q^2 \over s}} -
2 \langle {\bar q} q \rangle_\rho e^{-m_q^2/M^2}
\nonumber \\
&+& {8m_q \over M^2} \langle q^\dagger iD_0 q \rangle_\rho
e^{-m_q^2/M^2}
-{2m_q^2 \over M^2} \langle {\bar q} q \rangle_\rho e^{-m_q^2/M^2}\ ,
\label{pion1} \\
{{m^*_\pi}^2 \over 2 m_q} f_t f_s e^{-{m^*}^2_\pi/M^2}
&=&
{3 m_q \over 4 \pi^2} \int^{S_0}_{4m_q^2} d s e^{-s/M^2}
\sqrt {1 - {4 m_q^2 \over s}} -
2 \langle {\bar q} q \rangle_\rho e^{-m_q^2/M^2}
\nonumber \\
&-& {8 m_q \over 3 M^2 }
\langle q^\dagger iD_0 q \rangle_\rho e^{-m_q^2/M^2}
-{2m_q^2 \over M^2} \langle {\bar q} q \rangle_\rho e^{-m_q^2/M^2}
\nonumber \\
&+&{32 \over 3 M^2 } \left [ \langle {\bar q} (i D_0)^2 q \rangle_\rho
+ {1\over 8}
\langle {\bar q} g_s \sigma \cdot {\cal G} q \rangle_\rho \right ]
e^{-m_q^2/M^2}\ .
\label{pion2}
\end{eqnarray}
Here $S_0$ is the continuum threshold and $M$ is the Borel mass.
The various nuclear condensates appearing in the right-hand side
are evaluated in the linear density approximation, which yields
\begin{eqnarray}
&&\langle q^\dagger iD_0 q \rangle_\rho = \rho~ \langle q^\dagger
iD_0 q \rangle_N + {m_q \over 4} \langle {\bar q} q \rangle_\rho \
, \label{d4}
\\
&&\langle {\bar q} (iD_0)^2 q \rangle_\rho + {1\over 8} \langle
{\bar q} g_s \sigma \cdot {\cal G} q \rangle_\rho = \rho \left [
\langle {\bar q} (iD_0)^2 q \rangle_N + {1\over 8} \langle {\bar
q} g_s \sigma \cdot {\cal G} q \rangle_N \right ] + {m_q^2 \over 4
} \langle {\bar q} q \rangle_\rho\ , \label{d5}
\\
&&\langle {\bar q} q \rangle_\rho = \langle {\bar q} q \rangle_0 +
\rho~ \langle {\bar q} q \rangle_N\ . \label{d3}
\end{eqnarray}
Here the subscript ``$0$'' (``$N$'') denotes the vacuum (nucleon)
expectation value.  It is interesting to note that, when
Eqs.(\ref{d4}) and (\ref{d5}) are put into our sum rules, the
$m_q^2 \langle {\bar q} q \rangle_\rho$ terms are canceled away.

To leading order in quark mass, the ratio of the two sum rules gives
\begin{eqnarray}
{f_s \over f_t}
\simeq 1- {16 \over 3 M^2}~ {\langle {\bar q} (i D_0)^2 q \rangle_{\rho}
+ {1\over 8}
\langle {\bar q} g_s \sigma \cdot {\cal G} q \rangle_{\rho} \over
\langle {\bar q}q \rangle_\rho }\ .
\label{ratio}
\end{eqnarray}
Thus, the main splitting between the
two decay constants is driven
by the {\it nuclear} dimension 5 condensate,
\begin{eqnarray}
\langle {\bar q} (i D_0)^2 q \rangle_\rho
+ {1\over 8}
\langle {\bar q} g_s \sigma \cdot {\cal G} q \rangle_\rho\ .
\end{eqnarray}
Note, in vacuum, this dimension 5 condensate is zero and
both sum rules reproduce, up to leading order in quark mass,
the well-known Gell-Mann$-$Oakes$-$Renner (GOR) relation
(in vacuum, $f_t=f_s=f_\pi=131$ MeV),
\begin{eqnarray}
m_\pi^2 f_\pi^2=
-4 m_q \langle {\bar q} q \rangle_0 \ .
\label{gor}
\end{eqnarray}
Even including higher orders in quark mass, our sum rules
reproduce the vacuum sum rule for the pseudoscalar
decay constants~\cite{Kim:2001ss}.
Moreover the $\Pi_t$ sum rule, Eq.(\ref{pion1}), satisfies the in-medium
GOR relation
\begin{eqnarray}
{m_\pi^*}^2 f_t^2=
-4 m_q \langle {\bar q} q \rangle_\rho \ ,
\label{mgor}
\end{eqnarray}
to leading order in $m_q$.

\section{The dimension 5 condensate}
\label{sec:sign}

The dimension 5 condensate is crucial for splitting $f_t$ and $f_s$.
Depending on its sign,
we clearly have different prediction on the ratio $f_s/f_t$.
The ratio becomes greater (smaller) than the unity if the dimension 5
condensate is positive (negative).
Using $\langle {\bar q} g_s \sigma \cdot {\cal G} q \rangle
=2 \langle {\bar q} D^2 q \rangle $ in the chiral limit,
the dimension 5 condensate in the linear density approximation
can be rearranged into the form,
\begin{eqnarray}
\langle {\bar q} (i D_0)^2 q \rangle_\rho
+ {1\over 8}
\langle {\bar q} g_s \sigma \cdot {\cal G} q \rangle_\rho
=
\rho \left [ {3 \over 4} \langle {\bar q} (i D_0)^2 q \rangle_N
+ {1 \over 4} \langle {\bar q} (i{\bf D})^2 q \rangle_N
\right ] .
\end{eqnarray}
Ref.~\cite{Kim:2001xu} argued that, since $iD_0$ and $i{\bf D}$
are Hermitian operators, their square must be positive definite.
Thus, the nucleon dimension 5 condensate, ${3 \over 4} \langle
{\bar q} (i D_0)^2 q \rangle_N + {1 \over 4} \langle {\bar q}
(i{\bf D})^2 q \rangle_N$, is claimed to have the same sign with
the positive quantity,
$\langle {\bar q} q \rangle_N$,
which seems consistent with its rough estimate from the
bag model~\cite{Jin:1992id}.  This Hermitian argument however is not
consistent with the vanishing dimension 5 condensate {\it in vacuum}
\footnote{ If the same Hermitian argument is applied to its
vacuum expectation value, we should have
$ {3 \over 4} \langle {\bar q}
(i D_0)^2 q \rangle_0 + {1 \over 4} \langle {\bar q} (i{\bf D})^2
q \rangle_0\le 0$ as $\langle {\bar q} q \rangle_0 < 0 $.
Since the vacuum expectation value is zero,
the Hermitian argument only leads to a trivial
consequence in vacuum, $\langle {\bar q} (i D_0)^2 q \rangle_0
=\langle {\bar q} (i{\bf D})^2 q \rangle_0 =0$. This however can
not be correct because their difference is well-known to be
nonzero, $\langle {\bar q} (D_0^2-{\bf D}^2) q \rangle_0 =\langle
{\bar q} D^2 q \rangle_0 \ne 0$.}.

A possible resolution for this inconsistency can be sought for
by considering the condensate in
the Euclidean space, $x_0\rightarrow ix_4$.
In this space, the vacuum expectation value
becomes $ -{3 \over 4} \langle {\bar q} (i D_4)^2 q
\rangle_0 + {1 \over 4} \langle {\bar q} (i{\bf D})^2 q \rangle_0$
and, due to the $O(4)$ symmetry of vacuum, it is clear that the
two terms are canceled to make their sum zero,
agreeing with the expectation.
Its nucleon expectation value,
$ -{3 \over 4} \langle {\bar q} (i D_4)^2 q \rangle_N
+ {1 \over 4} \langle {\bar q} (i{\bf D})^2 q \rangle_N$, however, is not
necessarily zero because the $O(4)$ symmetry is broken by a
preferred direction of the nucleon state.
Now, because of the opposite
sign, the Hermitian argument does not lead to a definite sign for
the nucleon dimension 5 condensate.

We now comment on a rough estimate of the dimension 5 condensate using
the bag model~\cite{Jin:1992id}.  According to this,
$\langle {\bar q} (i D_0)^2 q \rangle_N$ is zero and,
depending on how one treats
$\langle {\bar q} g_s \sigma \cdot {\cal G} q \rangle_N$,
the dimension 5 condensate varies from
$0.08$ GeV$^2$ to $0.3$ GeV$^2$.
Though not precise, the bag model seems
to suggest a positive value for the dimension 5 condensate.
However, in the bag model estimate, the non-Abelian nature of
gluon fields can not be implemented and the quark field equation
is violated on the bag surface~\cite{Jin:1992id}.
Because of this limitation, in the subsequent analysis of the in-medium
nucleon sum rule~\cite{Jin:1993up}, the dimension 5
condensate is allowed to be negative,
varying from $-0.5$ GeV$^2$ to 0.5 GeV$^2$. Therefore,
the positive value for the dimension 5 condensate
as well as its magnitude is not firmly established.
One needs an alternative method to constrain this condensate.
In the following analysis, we look for this dimension 5 condensate
within our sum rules using the in-medium pion mass as an input.

\section{The nucleon intermediate state}
\label{sec:nucleon}

Another motivation for doing this update is to investigate more closely
the hadronic content of the correlation function
Eq.(\ref{corr}). In particular, the correlation function may pick
up some contributions from $N$ and $\Delta$ intermediate states
and the quasi-pion dominance has to be checked further. The
nuclear correlation function, Eq.(\ref{corr}), in the linear
density approximation is separated into the vacuum and nucleon
parts,
\begin{eqnarray}
\Pi^\mu = \Pi^\mu_0 + \rho \Pi^\mu_N
\label{lda_c}
\end{eqnarray}
where the nucleon correlation function is given by
\begin{eqnarray}
\Pi_N^\mu \equiv i \int d^4x~ e^{ip\cdot x} \langle N| T[{\bar d}(x) i
\gamma_5 u (x)~ {\bar u}(0) \gamma^\mu \gamma_5 d (0)] | N \rangle\ .
\label{ncorr}
\end{eqnarray}
The nucleon intermediate state
can make a pole structure at $p_0^2=0$, which may affect the sum
rules through Eq.(\ref{ncorr}) though its contribution
is down by the nuclear density $\rho$.

To calculate the $N$ intermediate state contribution, we
insert\cite{Christos:tu}
\begin{eqnarray}
\int {d^4 q\over (2\pi)^4}
\delta(q^2-m_N^2) \theta (q_0)~ |N({\bf q})\rangle \langle N({\bf q}) |
\end{eqnarray}
between the pseudoscalar and axial-vector currents in
Eq.(\ref{ncorr}). The nucleon matrix elements of the axial-vector
current and the pseudoscalar current are given respectively
by~\cite{Donoghue:1992dd}
\begin{eqnarray}
\langle N({\bf q})| {\bar d}\gamma_\mu \gamma_5 u | N({\bf k})\rangle
&=&{\bar U}_N ({\bf q}) \left [
g_A \gamma_\mu \gamma_5 + (q-k)_\mu \gamma_5 g_p \right ]
U_N({\bf k}) \\
\langle N({\bf q})| {\bar d}i\gamma_5 u | N({\bf k})\rangle
&=&-{f_\pi m_\pi^2 \over 2m_q}
{g_{\pi NN} \over m_\pi^2 -(q-k)^2}
{\bar U}_N ({\bf q}) i\gamma_5 U_N({\bf k})
\end{eqnarray}
where $U_N$ is the nucleon spinor, $g_A$ the nucleon
axial charge and $g_p$ the induced pseudoscalar term.
Inserting these into the nucleon correlation function, Eq.(\ref{ncorr}),
we directly evaluate the $N$ intermediate state contribution.
We find that this contribution is
proportional to
$\sqrt{{\bf p}^2+m_N^2}-m_N$, which vanishes in the limit of Eq.~(\ref{comb}).
Therefore,  the nucleon intermediate state does not contribute to the
correlation functions, $\Pi_t$ and $\Pi_s$.

\section{The $\Delta$ intermediate state}
\label{sec:delta}

We now consider the $\Delta$ contribution in the nucleon correlation
function, Eq.(\ref{ncorr}).
A $\Delta$ couples to $\pi N$ channel strongly
and the correlation function may pick up
a significant contribution from the $\Delta$ intermediate state.
The $\Delta$ intermediate state in $\Pi^\mu_N$
must have a pole at $p_0^2=(m_\Delta-m_N)^2 \sim
0.09$ GeV$^2$.  One can either treat this as a part of the continuum or
directly calculate this contribution using an effective model.
Unlike the nucleon intermediate case, however, the matrix elements involved,
$\langle \Delta | {\bar u} \gamma_\mu
\gamma_5 d | N \rangle$ and
$\langle \Delta | {\bar d}i\gamma_5 u | N \rangle$,
are not well-known.
Even if one can establish a form of the $\Delta$ contribution,
the parameters involved  highly depend on various models:
the estimate of this contribution can not be precise.

Roughly, one may eliminate the $\Delta$ contribution by considering the
sum rules with an additional weight of $s-(m_\Delta -m_N)^2$,
\begin{eqnarray}
\int^{S_0}_0 ds~ e^{-s/M^2} [s-(m_\Delta -m_N)^2]
{1 \over \pi} {\rm Im}
[\Pi_l^{phen} (s) - \Pi_l^{ope} (s)]=0  ~~~~(l=t,s) \ .
\label{deltas}
\end{eqnarray}
The new weight eliminates the pole at $p_0^2=(m_\Delta-m_N)^2$ and
the sum rules are now free from the $\Delta$ contribution.
However, the new weight affects the vacuum part of the sum
rules as well and, in the limit of
$\rho \rightarrow 0$,  the right-hand side of the GOR relation,
Eq.(\ref{gor}), entails the factor
\begin{eqnarray}
{m_q^2-(m_\Delta -m_N)^2 \over m_\pi^2-(m_\Delta
-m_N)^2} \sim 1.3
\end{eqnarray}
coming from the additional weight.
Though this factor is the unity to leading order in the chiral expansion,
its numerical value deviated from the unity affects the GOR relation.
Thus, this way of eliminating the $\Delta$ contribution
is crude.

A more economical way is to
apply the similar prescription only to the {\it nucleon} correlation
function $\Pi_N^\mu$ in Eq.(\ref{lda_c}). That is, we introduce
$\Pi_{Nt}$ and $\Pi_{Ns}$ from $\Pi_N^\mu$ similarly defined
as $\Pi_t$ and $\Pi_s$ in Eq.(\ref{comb})
and construct
\begin{eqnarray}
\int^{S_0}_0 ds~ e^{-s/M^2} [s-(m_\Delta -m_N)^2]
{1 \over \pi} {\rm Im}
[\Pi_{Nl}^{phen} (s) - \Pi_{Nl}^{ope} (s)]=0  ~~~~(l=t,s) \ .
\label{deltasum}
\end{eqnarray}
Obviously, this prescription does not suffer from the
problem mentioned above: it does not affect the GOR relation in vacuum. To
construct $\Pi_{Nt}^{phen}$ and $\Pi_{Ns}^{phen}$, we expand in terms
of the density the in-medium parameters appearing in Eq.(\ref{phen}),
\begin{eqnarray} f_t = f_\pi + \rho \Delta f_t \;;
\quad \quad f_s = f_\pi + \rho \Delta f_s \;; \quad \quad m_\pi^*
= m_\pi + \rho \Delta m_\pi\ . \label{deltapar}
\end{eqnarray}
The new phenomenological parameters, $\Delta f_t$, $\Delta f_s$
and $\Delta m_\pi$, constitute $\Pi_{Nt}^{phen}$ and $\Pi_{Ns}^{phen}$.
In the OPE side, using Eqs.(\ref{d4}),(\ref{d5}) and (\ref{d3}) for
various nuclear condensates, we collect
the terms proportional to $\rho$ corresponding
to $\Pi_{Nt}^{ope}$ and $\Pi_{Ns}^{ope}$.
We put them into Eq.(\ref{deltasum}) and obtain the $\Delta$
subtracted sum rules for $\Delta f_t$ and $\Delta f_s$,
\begin{eqnarray}
\Delta f_t
&=& -{\Delta m_\pi \over m_\pi} f_\pi
-{2m_q r_\Delta \over m_\pi^2 f_\pi} \langle {\bar q} q \rangle_N
e^{(m_\pi^2-m_q^2)/M^2}
\nonumber \\
&+& {8 m_q^2 \over m_\pi^2 f_\pi}  \langle q^\dagger iD_0 q
\rangle_N \left [ {r_\Delta \over M^2}- \pi_\Delta \right ]
e^{(m_\pi^2-m_q^2)/M^2}\ , \label{deltasum1}
\\
\Delta f_t+ \Delta f_s
&=&  -{2 \Delta m_\pi \over m_\pi} f_\pi
-{4m_q r_\Delta \over m_\pi^2 f_\pi} \langle {\bar q} q \rangle_N
e^{(m_\pi^2-m_q^2)/M^2}
\nonumber \\
&-& {16 m_q^2 \over 3 m_\pi^2 f_\pi } \langle q^\dagger iD_0 q \rangle_N
\left [ {r_\Delta \over M^2} -\pi_\Delta
\right ] e^{(m_\pi^2-m_q^2)/M^2}
\nonumber \\
&+& {64 m_q \over 3 m_\pi^2 f_\pi } \left [ \langle {\bar q} (i
D_0)^2 q \rangle_N + {1\over 8} \langle {\bar q} g_s \sigma \cdot
{\cal G} q \rangle_N \right ] \left [ {r_\Delta \over
M^2}-\pi_\Delta \right ] e^{(m_\pi^2-m_q^2)/M^2}\ ,
\label{deltasum2}
\end{eqnarray}
where we have defined
\begin{eqnarray}
r_\Delta = {m_q^2-(m_\Delta -m_N)^2 \over m_\pi^2-(m_\Delta
-m_N)^2}  \;; \quad \quad
\pi_\Delta = {1 \over m_\pi^2-(m_\Delta -m_N)^2}\ .
\end{eqnarray}
Note, the pion mass shift in the exponential, which is an
order ${\cal O} (m_\pi^3)$ or higher, has been neglected in deriving
these sum rules. Once $\Delta m_\pi$ is given, these sum rules yield
$\Delta f_t$ and $\Delta f_s$, which then, through
Eq.(\ref{deltapar}), lead to $f_t$ and $f_s$ with the $\Delta$
contribution being subtracted. For the justification of
this, we have checked that, without
$\Delta$ subtraction, this procedure gives $f_t$ and $f_s$ similar
in magnitude with those obtained directly from
Eqs.(\ref{pion1}),(\ref{pion2}).

One may worry about the $\Delta$ decay width
and argue that the $\Delta$ contribution is not a pole.
A $\Delta$ in free space strongly decays to $\pi N$ with its width
$115$ MeV. However, in nuclear matter, a $\Delta$ {\it at rest} can not
decay to $\pi N$ by the Pauli blocking~\footnote{
Note, the limit of Eq.(\ref{comb}) means that we are considering
a $\Delta$ at rest.}.
On the other hand, a $\Delta$ in nuclear matter can have
a ``spreading width'' through the mechanism
$\Delta + N \rightarrow N + N$.
Its width at nuclear saturation density is between
57 to 75 MeV~\cite{Kim:1996ad,Jain:kf}, reasonably small.
The pole ansantz for a resonance with this small width
is believed to be reasonable.

\section{Sum Rule Analysis}
\label{sec:analysis}

As we have discussed, the dimension 5 condensate
is important for splitting $f_s$ and $f_t$ but its value is not well-known.
An additional information is necessary to restrict
the value of the dimension 5 condensate. Instead of relying
on a model calculation, we look for its value directly from our sum rules
using in-medium pion mass as an input.
The in-medium pion mass has been studied extensively by chiral perturbation
theory~\cite{Meissner:2001gz,Park:2001ht,Delorme:1996cc,Kaiser:2001bx}.
Experimentally, it is extracted from the local potential of the deeply
bound pionic atom in $^{208}$Pb~\cite{Yamazaki:1997na}.
A consensus from these studies
is that the in-medium pion mass increases slightly up to 20 MeV.
We therefore look for an optimal value of the dimension 5 condensate
that leads to the in-medium pion mass within $139 \sim 159$ MeV
from the $\Pi_s$ sum rule, Eq.(\ref{pion2}).
The $\Pi_t$ sum rule, Eq.(\ref{pion1}), can not be used for this purpose
as it does not depend on the dimension 5 condensate.

\begin{figure}
\caption{ The in-medium pion mass squared, obtained from
Eq.(\ref{pionmass}), is plotted with respect to the Borel mass at the
saturation density $\rho_0 =0.17$ fm$^{-3}$. The number indicated
in each curve is the value of the dimension 5 condensate, $D_5$,
used. The positive value of $D_5$ leads to the tachyonic mass,
${m_\pi^*}^2 < 0$. } \label{fig1}

\setlength{\textwidth}{6.1in}   
\setlength{\textheight}{9.in}  
\centerline{%
\vbox to 2.4in{\vss
   \hbox to 3.3in{\includegraphics{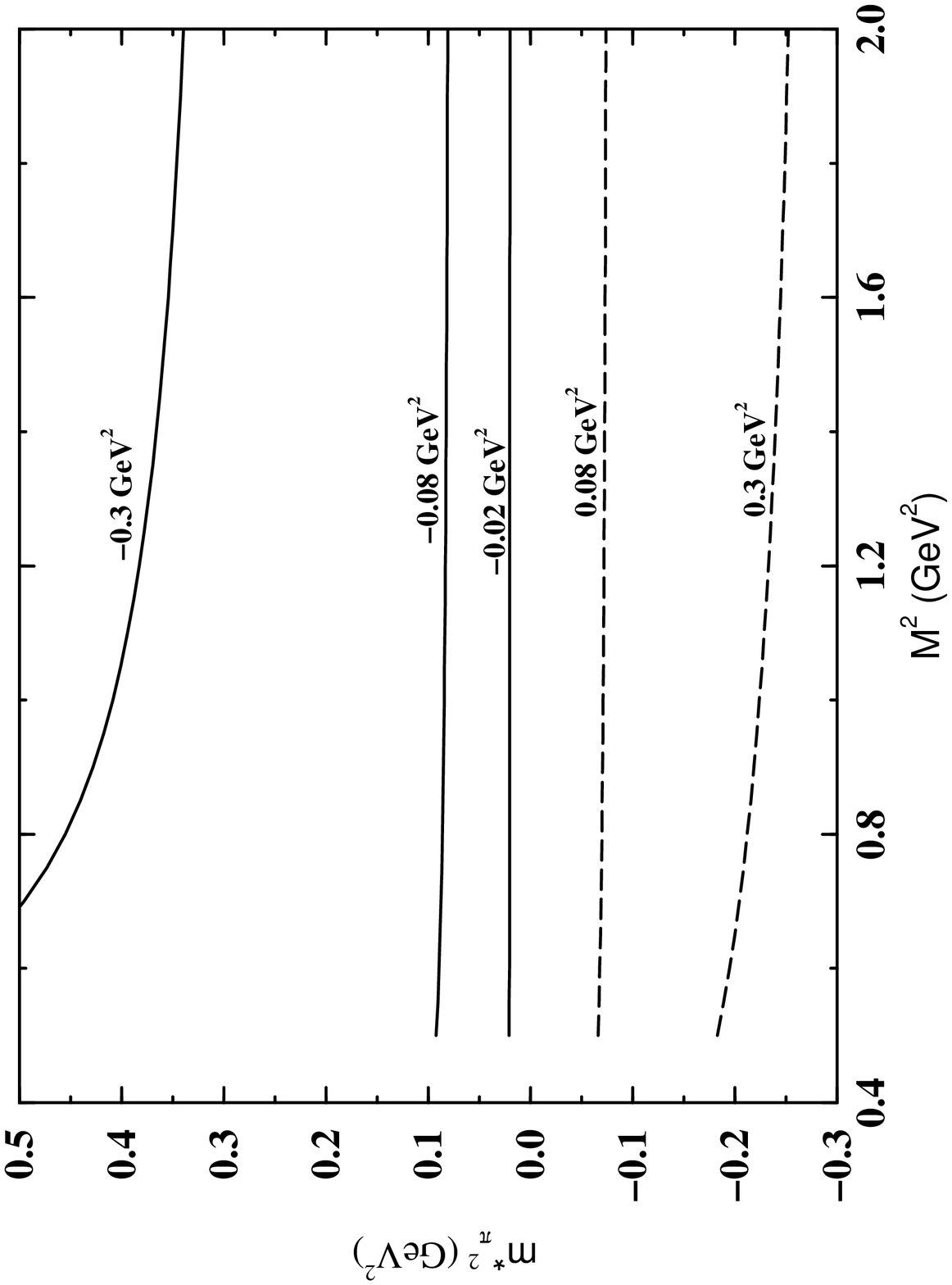}\hss}}
}
\vspace{100pt}
\end{figure}

To calculate the pion mass from our sum rules,
we take the derivative of Eq.(\ref{pion2}) with respect to $1/M^2$
and divide the resulting formula by Eq.(\ref{pion2}).
Namely, by defining the right-hand side of Eq.(\ref{pion2})
by $\Pi_{Borel} (M^2)$
and its derivative with respect to $1/M^2$ by $\Pi_{Borel}^\prime (M^2)$,
the in-medium pion mass satisfies
\begin{eqnarray}
-{m_\pi^*}^2 = {\Pi_{Borel}^\prime (M^2)
\over \Pi_{Borel} (M^2) }\ .\label{pionmass}
\end{eqnarray}
Using this formula, we plot ${m_\pi^*}^2$ versus the Borel
mass $M^2$ in fig.~\ref{fig1} for various values of
the nucleon dimension 5 condensate,
\begin{eqnarray}
D_5 \equiv \langle {\bar q} (i D_0)^2 q \rangle_N
+ {1\over 8}
\langle {\bar q} g_s \sigma \cdot {\cal G} q \rangle_N\ .
\end{eqnarray}
The continuum threshold is set to be $s_0=0.09$ GeV$^2$
corresponding to the $\Delta$ intermediate state, {\it i.e.,}
$(m_\Delta-m_N)^2$ but the result is not sensitive to this choice.
Other parameters used in our analysis are
\begin{eqnarray}
&&m_q = 0.007~{\rm GeV}
\;; \quad \quad
\langle {\bar q} q \rangle_0 = (-0.225~{\rm GeV})^3 \nonumber \\
&&\langle {\bar q} q \rangle_N = {0.045~{\rm GeV} \over 2 m_q}
\;; \quad \quad
\langle q^\dagger iD_0 q \rangle_N=0.18 ~{\rm GeV}\ .
\end{eqnarray}
As shown, we have quite different curves
depending on the dimension 5 condensate, $D_5$.
In particular, the positive value of $D_5$
leads to a tachyonic pion mass, ${m_\pi^*}^2 < 0$.
As the positive value
leads to $f_s/f_t \ge 1$ according to Eq.(\ref{ratio}),
the tachyonic pion mass for $D_5 \ge 0$ is consistent
with the claim that $f_s/f_t \ge 1$ breaks the
causality~\cite{Pisarski:1996mt}.
When $D_5$ is fixed to be around $-0.02$ GeV$^2$, the
pion effective mass is about 140 MeV and the stability is quite good
over a wide range of the Borel masses.
For the in-medium pion mass within
$139~{\rm MeV} \le m_\pi^* \le 159~{\rm MeV}$, $D_5$ is
{\it negative} and its magnitude is
within the range, $0.019 \le | D_5 | \le 0.025 $.
We have also checked that this range of $D_5$ is stable
under the rough subtraction of the $\Delta$ contribution
given in Eq.(\ref{deltas}).  Note, the $\Delta$ subtraction
procedure advocated in Eq.(\ref{deltasum}) is not
applicable for obtaining the in-medium pion mass.

\begin{figure}
\caption{The Borel curves for the ratio $f_t/f_\pi$
are plotted at the nuclear
saturation density $\rho_0 =0.17$ fm$^{-3}$. The solid lines are
obtained from the $\Pi_t$ sum rule, Eq.(\ref{pion1}), and the dashed
lines are when the $\Delta$ contributions are subtracted according
to Eqs.(\ref{deltapar}) and (\ref{deltasum1}).
The value of $m_\pi^*$ used to obtain these curves is
indicated.} \label{fig2}

\setlength{\textwidth}{6.1in}   
\setlength{\textheight}{9.in}  
\centerline{%
\vbox to 2.4in{\vss
   \hbox to 3.3in{\includegraphics{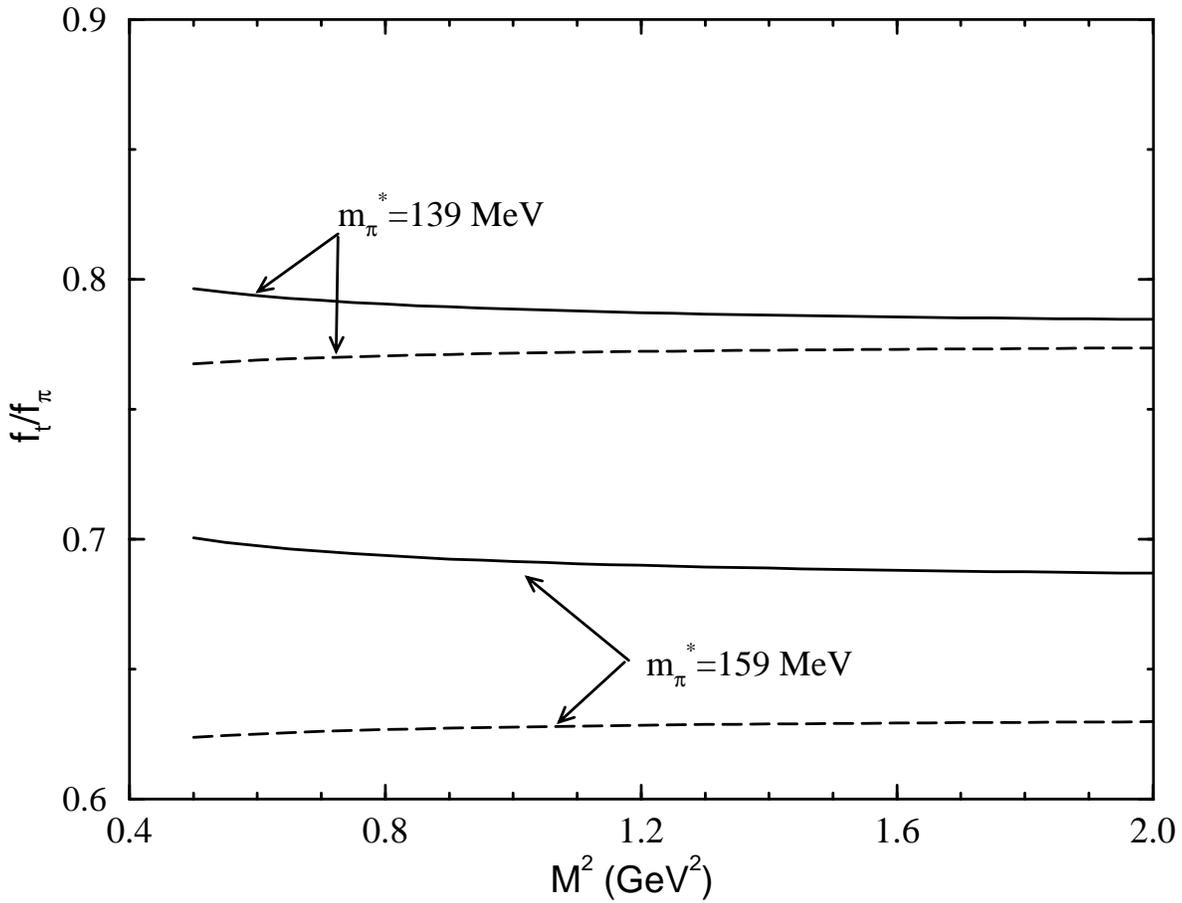}\hss}}
}
\vspace{100pt}
\end{figure}

We now move to an analysis for the pion decay constants, $f_t$ and $f_s$.
In fig.\ref{fig2}, we plot $f_t/f_\pi$ using the sum rule Eq.(\ref{pion1})
at the saturation density $\rho =0.17$ fm$^{-3}$.
When $m_\pi^*=139$ MeV, $f_t/f_\pi = 0.79$ is obtained from
the $\Pi_t$ sum rule Eq.(\ref{pion1}) at $M^2=1$ GeV$^2$ but with
$\Delta$ subtracted according to Eqs.(\ref{deltapar}) and (\ref{deltasum1})
the ratio becomes slightly smaller, 0.77.
The Borel curves for these two cases are shown by the
two upper curves in fig.~\ref{fig2}.  The solid curve is from
Eq.(\ref{pion1}) and the dashed curve is when the $\Delta$
contribution is subtracted.
The lower two curves are obtained when we use the larger pion mass,
$m_\pi^*=159$ MeV.
Increasing the pion mass makes $f_t/f_\pi$ smaller in both cases.
Because the $\Pi_t$ sum rule
satisfies the in-medium GOR relation, Eq.(\ref{mgor}),
large $m_\pi^*$ is compensated by small $f_t$.
Also at $m_\pi^* = 159$ MeV, $f_t$ from the $\Delta$ subtracted sum rule
is  9 \% smaller than the one without $\Delta$ subtraction.
Large $m_\pi^*$
may easily excite a $\Delta$ so that
the $\Delta$ contribution becomes larger in the correlation
function.
It is interesting to note that $f_t/f_\pi$ at $m_\pi^*=139$ MeV,
either with or without $\Delta$ subtraction,
is not far from the experimental value of 0.8~\cite{Suzuki:2002ae}.

\begin{figure}
\caption{The Borel curves for the ratio $f_s/f_\pi$
are plotted at the nuclear
saturation density $\rho_0 =0.17$ fm$^{-3}$. The solid lines are
obtained from Eq.(\ref{pion2}). The dashed
lines are when the $\Delta$ contributions are subtracted according
to Eqs.(\ref{deltapar}) and (\ref{deltasum2}). } \label{fig3}

\setlength{\textwidth}{6.1in}   
\setlength{\textheight}{9.in}  
\centerline{%
\vbox to 2.4in{\vss
   \hbox to 3.3in{\includegraphics{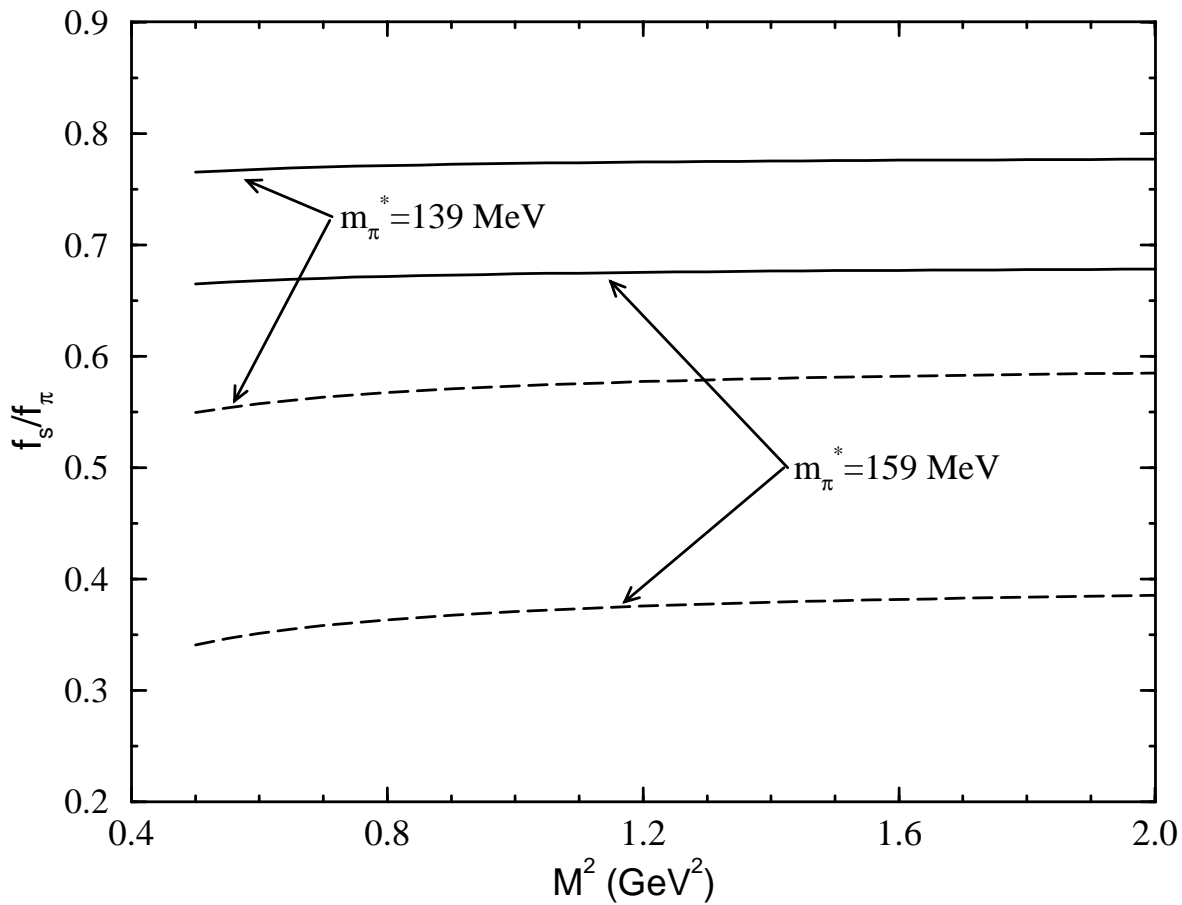}\hss}}
}
\vspace{100pt}
\eject
\end{figure}

Fig.~\ref{fig3} shows the ratio $f_s/f_\pi$.
To obtain this, we first calculate $f_s/f_t$ from the
ratio of Eqs.(\ref{pion1}) and (\ref{pion2}), and
multiply it by $f_t/f_\pi$ obtained from fig.~\ref{fig2}.
The results with $\Delta$ subtraction from
Eqs.(\ref{deltasum1}) and (\ref{deltasum2}) are shown
by the dashed curves. Unlike to the $f_t/f_\pi$ case,
a somewhat sizable suppression of
$f_s/f_\pi$ is obtained when the $\Delta$ contribution is subtracted.
This shows that the $\Pi_s$ sum rule depends heavily on
the $\Delta$ contribution.
At $m_\pi^*=139$ MeV, the $\Delta$ subtracted sum rule
shifts $f_s/f_\pi$ from 0.78 to 0.57, 27\% change, while
at $m_\pi^*=159$ MeV, from 0.68 to 0.37, 45 \% change.
Such a large suppression of $f_s$ is similar to the results from
the in-medium chiral perturbation
theory~\cite{Kirchbach:1997rk,Meissner:2001gz}.
These results as well as the results for  $f_t$
are summarized in Table~\ref{tab1}.

\begin{figure}
\caption{The density dependence of the ratios $f_t/f_\pi$
calculated at $m_\pi^*=139$ MeV and the Borel mass
$M^2=1$ GeV$^2$.  The solid line (the dashed line)
is obtained without (with) $\Delta$ subtraction.
} \label{fig4}

\setlength{\textwidth}{6.1in}   
\setlength{\textheight}{9.in}  
\centerline{%
\vbox to 2.4in{\vss
   \hbox to 3.3in{\includegraphics{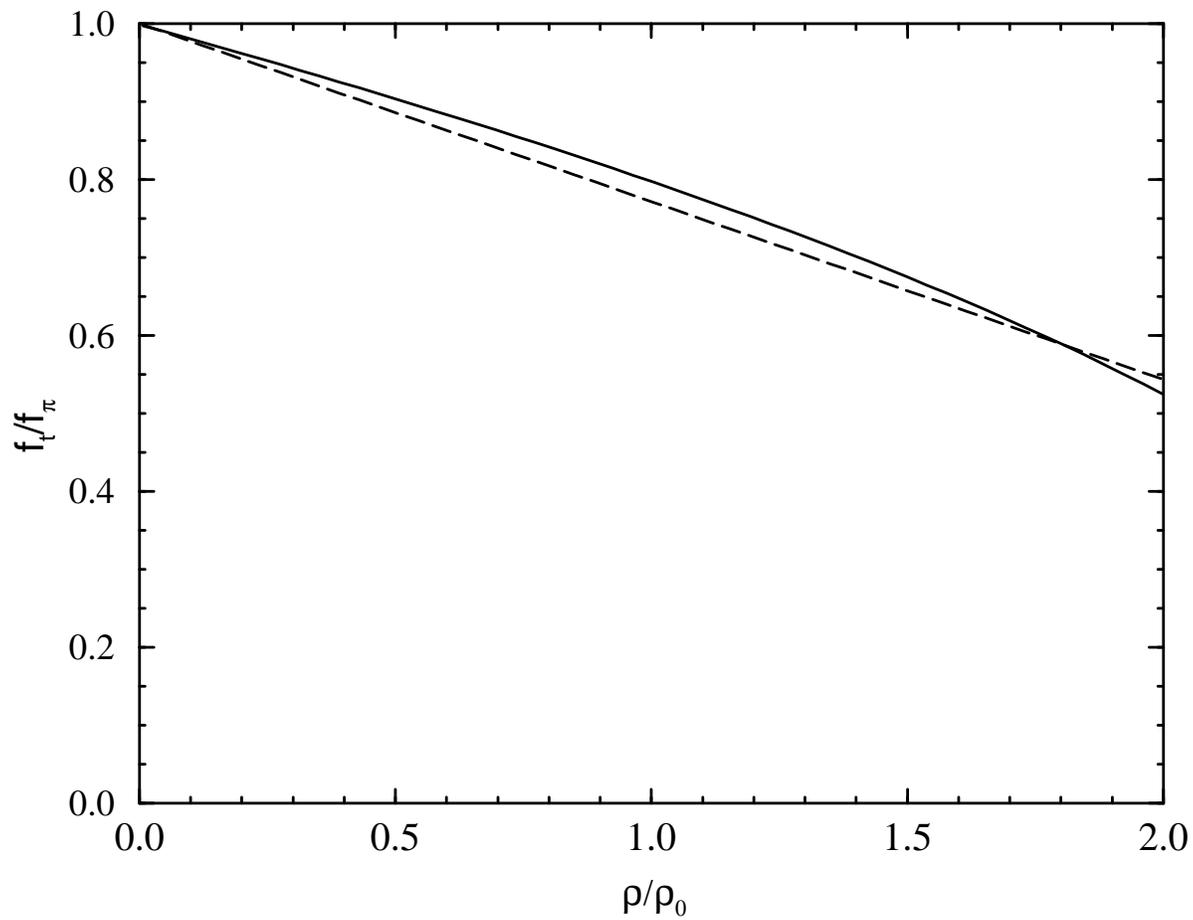}\hss}}
}
\vspace{100pt} \eject
\end{figure}

\begin{figure}
\caption{The density dependence of the ratios
$f_s/f_\pi$ calculated at $m_\pi^*=139$ MeV and the Borel mass
$M^2=1$ GeV$^2$. The solid line (the dashed line)
is obtained without (with) $\Delta$ subtraction.
This shows that $f_s/f_\pi$ has a
strong dependence on the $\Delta$ contribution. } \label{fig5}

\setlength{\textwidth}{6.1in}   
\setlength{\textheight}{9.in}  
\centerline{%
\vbox to 2.4in{\vss
   \hbox to 3.3in{\includegraphics{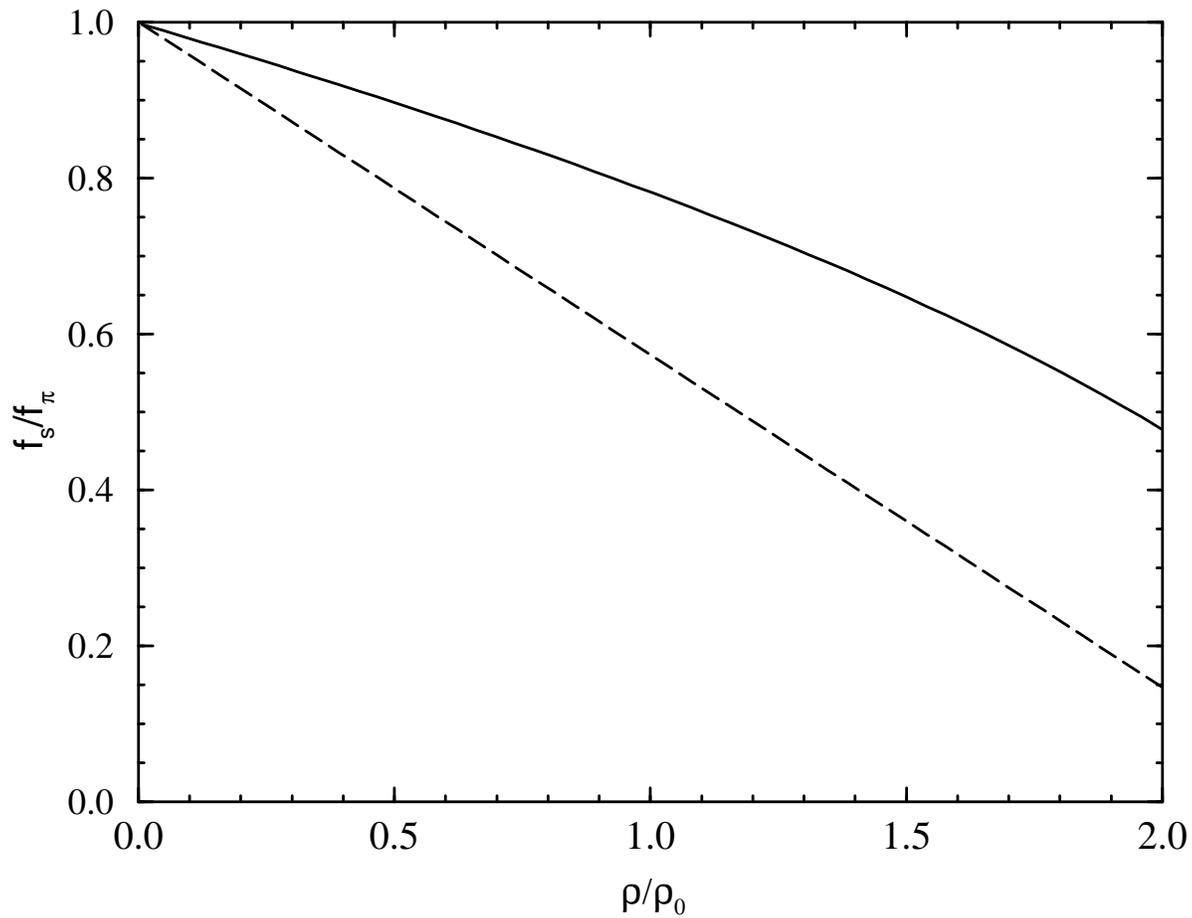}\hss}}
}
\vspace{100pt} \eject
\end{figure}

Fig.~\ref{fig4} shows the density dependence
of $f_t/f_\pi$. The solid curve is obtained from
Eq.(\ref{pion1}) with $m_\pi^*=139$ MeV and the dashed
curve is when the $\Delta$ contribution is subtracted
according to Eqs.(\ref{deltapar}) and (\ref{deltasum1}).
The similar curves for $f_s/f_\pi$ are shown in Fig.~\ref{fig5}.
Again, we see that $f_s/f_\pi$ is strongly suppressed when
the $\Delta$ contribution is subtracted.
It should be also noted that $f_s/f_t$
is less than the unity always, agreeing with
the causal constraint derived from Ref.~\cite{Pisarski:1996mt}.

\begin{table}
\caption{The summary table for the decay constants,
$f_t/f_\pi$ and $f_s/f_\pi$ (with $f_\pi=131$ MeV)
calculated from our sum rules for given in-medium pion masses.
The $D_5$ value is obtained from Eq.(\ref{pionmass}).
The numbers in parenthesis are when the $\Delta$ contribution is
subtracted according to Eq.(\ref{deltasum}). }

\begin{center}
\begin{tabular}{cccc}
$m_\pi^*$ (MeV) & $D_5$ (GeV$^2$) & $f_t/f_\pi$ & $f_s/f_\pi $  \\
\hline\hline
139 & -0.019 & 0.79 (0.77) & 0.78 (0.57)       \\
159 &-0.025 & 0.69 (0.63) &   0.68 (0.37)   \\
\hline\hline
\end{tabular}
\end{center}
\label{tab1}

\vspace{50pt}

\end{table}

\section{Summary and outlook}
\label{sec:summary}

We have updated the QCD sum rule calculation of the in-medium
pion decay constants, $f_t$ and $f_s$, using pseudoscalar and
axial vector correlation function.
We have argued that the dimension 5 condensate, which is
crucial for splitting between $f_t$ and $f_s$, is not
necessarily restricted to be positive.  In fact,
the pion mass calculated from the $f_t$ sum rule takes a real
value  when the dimension 5 condensate is restricted to be negative.
We have taken into account contributions from the $N$ and $\Delta$
intermediate states in the correlation function.
The $N$ intermediate state was found not to contribute
to our sum rules. For the $\Delta$ contribution,
we have included either in the continuum or explicitly eliminated
by putting an additional weight in the sum rules.
The $\Delta$ subtraction procedure was found to affect the extraction of
$f_t/f_\pi$ slightly by $3 -9 \%$.
For $f_s/f_\pi$, it affects  strongly
by $27 -45 \% $. This strong suppression of $f_s$ is
similar to the results from the in-medium chiral perturbation theory.
In future, it will be interesting to apply our method for kaonic
channel and investigate the in-medium $f_K$.
As the strange quark mass is not small,
the gluonic dimension 5 operator $m_s \left \langle {\cal G}^2
\right \rangle$, which is absent in this work, could be important.

\acknowledgments
The work by Hungchong Kim was supported by the
Korea Research Foundation Grant KRF-2002-015-CP0074.

\eject

\end{document}